# On the Origin of the Second-Order Nonlinearity in Strained Si-SiN Structures

J. B. Khurgin,[1*] T. H. Stievater,[2] M. W. Pruessner,[2] W. S. Rabinovich[2]

[1]*Johns Hopkins University, Baltimore, MD 21218*
[2]*Naval Research Laboratory, Washington, DC 20375*
*Corresponding author jakek@jhu.edu*

**The development of efficient low-loss electro-optic and nonlinear components based on silicon or its related compounds, such as nitrides and oxides, is expected to dramatically enhance silicon photonics by eliminating the need for non-CMOS-compatible materials. While bulk Si is centrosymmetric and thus displays no second-order ($\chi^{(2)}$) effects, a body of experimental evidence accumulated in the last decade demonstrates that when a strain gradient is present, a significant $\chi^{(2)}$ and Pockels coefficient can be observed. In this work we connect a strain-gradient-induced $\chi^{(2)}$ with another strain-gradient-induced phenomenon, the flexoelectric effect. We show that even in the presence of an extremely strong strain gradient, the degree by which a nonpolar material like Si can be altered cannot possibly explain the order of magnitude of observed $\chi^{(2)}$ phenomena. At the same time, in a polar material like SiN, each bond has a large nonlinear polarizability, so when the inversion symmetry is broken by a strain gradient, a small (few degrees) re-orientation of bonds can engender $\chi^{(2)}$ of the magnitude observed experimentally. It is our view therefore that the origin of the nonlinear and electro-optic effects in strained Si structures lies in not in the Si itself, but in the material providing the strain: the silicon nitride cladding.**

***OCIS codes:** (130.4310) Integrated Optics, nonlinear; (160.4330) Nonlinear optical materials; (160.2100) Electro-optical materials.*

## 1. INTRODUCTION

Silicon photonics [1] is a rapidly expanding field promising the monolithic integration of electronic and photonic devices into versatile optoelectronic circuits, including both passive and active optical components. Optical modulators are a key components of any photonic circuit and while a variety of Si modulators [2,3] have been demonstrated, their performance characteristics remain inferior to other modulators, such as LiNbO$_3$, GaAs, or polymers and other electro-optic materials in which the index change is achieved by means of Pockels effect. Unlike these compounds, Si is a centrosymmetric material, and the Pockels (or linear electro-optic) effect is absent in it. Therefore, a refractive index change in Si is usually achieved using carrier injection and depletion, which limits the speed, increases insertion loss, and reduces linearity of the modulation. When it comes to all-optical devices, such as frequency converters, parametric oscillators, and so on, once again the inversion symmetry of the Si lattice makes all the second-order (or $\chi^{(2)}$) effects vanish, requiring the use of less efficient third order (or $\chi^{(3)}$) phenomena. To achieve high performance one is often forced to use bonding techniques to integrate passive Si waveguides with active LiNbO$_3$, polymers [4] or III-V [5] components. The significance of a method to alter the crystal structure of Si to obtain substantial $\chi^{(2)}$ (and as a special case of it, a nonzero Pockels coefficient) could not be overestimated.

So, it is not surprising that when the linear electro-optic effect was first observed in strained Si [6] it was greeted with excitement, because not only was the effect strong, but also it could be engineered by changing the strain [7]. Prior works on second harmonic generation in Si [8,9] were also re-discovered around that time, and, while many researchers have concentrated on the realization of strained-silicon devices for integrated optics [10-11], a number of extensive studies with the goal of determining the cause of second order nonlinearity in Si has been performed [12-14]. In that respect, the inversion symmetry of Si could only be broken in the presence of an inhomogeneous strain [15]. This implies that the second order susceptibility is proportional to the strain gradient, which makes it much smaller than the value experimentally measured. A number of theoretical works [8,16] have attempted to explain the origin of the second order nonlinearity and Pockels effect, but none agreed with measurements. In the most recent work [17] it was suggested that interface charge effects can play an important role, which might partially account for the slow Pockels effect, but they cannot offer a plausible cause for fast second harmonic generation. Essentially, as pointed in [15], the Si-Si bond is homo-polar and has no second order hyper-polarizability [18], and hence $\chi^{(2)}$=0. Therefore, to attain a $\chi^{(2)}$ value that is on the order of tens of pm/V, as measured in [6,10,12-14] (i.e. comparable to that in III-V materials and perovskites, such as LiNbO$_3$), the strain gradient must induce fields that would produce asymmetric distribution of bond charge similar to that in heteropolar materials. Such fields then must be comparable in the magnitude to the intrinsic fields in the Si bonds themselves, i.e. roughly $10^8$-$10^9$ V/m, which, as a simple estimate made in [15] and expanded further on in the present paper clearly deems unfeasible.

But if the strain gradient cannot make the pure covalent bond polar enough to produce the measured second order effects, why can't it simply re-arrange the already polar bonds to produce a large χ(2), just as an electric field induces a non-zero χ(2) in electro-optic and nonlinear polymers? The polar bonds are already present in all the strained Si structures in which the second order nonlinear effects have been measured. These bonds belonged to the "stressors": the Si₃N₄ or SiO₂ cladding layers deposited onto Si. While the SiO₂ bond is almost entirely ionic and thus has a small bond charge, tetragonal Si-N bonds in silicon nitride are similar to those in GaAs – the charge is concentrated in the bond and is shifted towards N atoms. The four bonds comprising each tetragon do produce a large nonlinear polarizability, but, since silicon nitride is an amorphous, or, rather, a ceramic material, the tetragons are randomly oriented and these polarizabilities cancel. It is our hypothesis that a strain gradient can "orient" these tetragons in a preferential direction, resulting in a non-zero χ (2). This hypothesis is based on well-known recent results that show silicon nitride films are known to possess both ☐(2) and a Pockels coefficient [19-23] whose origin has not yet been fully understood. We also make clear the connection between the strain-gradient induced χ(2) (SGI-χ(2)) and flexo-electric effect [24-25], in support of our hypothesis. Our estimate shows that only "orientation" or "poling" of the Si-N bonds under the strain gradient can explain the order of magnitude of the SGI-χ(2) and SGI Pockels coefficients observed in all the experiments to-date.

## 2. ESTIMATE OF STRAIN-GRADIENT INDUCED X(2) In Si

The second order nonlinearity in heteropolar covalent materials originates from the asymmetry of the individual bonds between the anion and cation (e.g., a group III such as As and group V such as Ga ions). The effective potential difference between the anion and cation that pulls the bond charge towards the anion (As) is typically referred to as the heteropolar contribution to the bandgap C [26-28], which is commensurate with the difference in the electronegativities of the anion and cation. For GaAs this value equals 4.3 eV [29]. According to the bond charge theory of nonlinear susceptibilities [18], nonlinear polarizabilities of individual bonds are proportional to C, until C becomes very large and the bond turns ionic (This is what happens in crystalline SiO₂, quartz, where ☐is small despite a lack of inversion symmetry). The second order susceptibility ☐(2) then results from a summation over the individual bonds. For amorphous materials this summation yields zero, but for the stoichiometric arrangement of a zinc blende lattice this summation yields a large non-zero $\chi^{(2)}_{xyz}$ (the susceptibility is large along the directions of four tetragonal bonds). One can show that within an order of magnitude, the second order susceptibility can be estimated as $\chi^{(2)} \sim \chi^{(3)} E_{eff}$, i.e. the product of the third order susceptibility $\chi^{(3)}$ and the effective field caused by the potential difference between the anion and cation, $E_{eff} \approx C/ea_b$ where $a_b$ is the bond length. For GaAs $a_b = 2.4 \text{Å}$ and $\chi^{(3)} \approx 0.5 \times 10^{-19} m^2/V^2$, which yields $E_{eff} \approx 2 \times 10^{10} m/V$ and an estimated value of the induced second order nonlinearity of $\chi^{(2)}_{est} \approx 6 \times 10^{-10} m/V$ which is well within the order of magnitude from the actual value of $\chi^{(2)}_{GaAs} \approx 3.4 \times 10^{-10} m/V$. Similar results can be obtained for other materials with tetragonal bonding, like ZnSe and GaP [30]. Therefore, in general, the order of magnitude of SGI-☐(2) can be estimated for strained silicon as

$$\chi^{(2)}_{SGI} \sim \chi^{(3)}_{Si} E_{eff(\nabla u)} \quad (1)$$

where $E_{eff(\nabla u)}$ is the effective field produced by charge re-distribution due to a strain gradient.

The third order susceptibilities of Si and GaAs are comparable in magnitude (they differ by a factor of 2 as $\chi^{(3)}_{Si} \approx 10^{-19} m^2/V^2$ [31]). The reported value of SGI-χ(2) varies in magnitude, but there is a body of evidence [6,10,12-14]it they can be as high as a few tens of pm/V , or, roughly 10-20% of GaAs χ(2). To achieve this, an effective electric field on the scale of $10^9 V/m$ would be required At the same time, the intrinsic, or bonding field in Si can be estimated as roughly $E_b \sim e/4\pi\varepsilon_0 a_b^2 \approx 2.5 \times 10^{10} V/m$ (this field is also commensurate with the breakdown field of Si). Therefore, roughly a 4% distortion in the lattice over one bond length would be needed to attain the observed $\chi^{(2)}$ in strained Si, meaning that the strain gradient should be $\nabla u \approx 0.04/a_b \sim 10^8 m^{-1}$, roughly 3-4 orders of magnitude larger than in a typical strained Si.

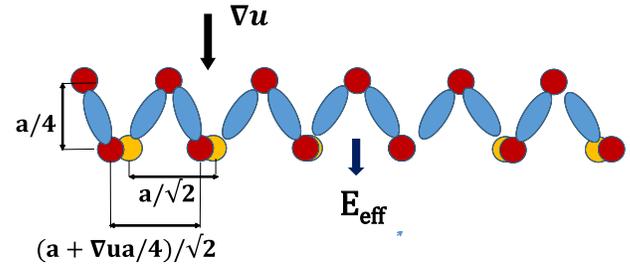

Fig. 1. Origin of the effective field acting upon Si bonds under strain gradient

To confirm this order-of-magnitude assessment we consider a simple model of how the strain gradient engenders the effective field acting on the bonds in Si. As shown in Fig.1, each bond is contained between two planes of Si atoms separated by the distance a/4, where a=5.6 Å is a lattice constant. Under the strain gradient $\nabla u$ the lattice constants in the two layers differ by $\delta a \approx (\nabla u \cdot a/4)a$ and the sheet density of Si atoms vary by $\delta N_{Si} = \delta(1/a^2) \approx \nabla u/2a$. The difference between the sheet charges in the two planes then produces the effective field $E_{eff(\nabla u)} \sim e\delta N_{Si} e/\varepsilon_0 = e\nabla u/2a\varepsilon_0$. Even assuming a 1% strain gradient over 1 micrometer ( $\nabla u = 10^4 m^{-1}$) one obtains $E_{eff(\nabla u)} \approx 1.6 \times 10^5 V/m$, sufficient to produce $\chi^{(2)}_{SGi} < 0.02 pm/V$, i.e. the aforementioned 3 to 4 orders of magnitude less than that measured in experiments.

Thus, simple motion of ions induced by the strain gradient is not sufficient to engender large second order nonlinearities. This can be generalized, of course, for any single crystal material. To further explore the effect of strain gradients, it is instructive to turn attention to another phenomenon associated with it – flexoelectric effect in solids.

## 3 FLEXOELECTRIC EFFECT AND ITS RELATION WITH STRAIN-GRADIENT INDUCED X[(2)]

The flexoelectric effect [24, 25] is an electromechanical effect in which the dielectric polarization P exhibits a linear response to a gradient of mechanical strain $\nabla u$. Long known to exist in liquid crystals and soft matter [32] it was first identified in solids theoretically by Mashkevich and Tolpygo [33, 34] and the first phenomenological framework for the description of this effect was offered by Kogan [35] in 1964, while the microscopic theory was developed soon afterword by Harris [36]. Since then, the flexoelectric effect has been observed in both ceramics [37, 38] and single crystals [39-41]. It was found to be significantly stronger in ceramics than in single crystals – a fact that that is highly relevant to the present work, as we establish the connection between flexoelectricity and SGI-$\chi^{(2)}$.

The flexoelectric effect is usually introduced [24] by generalizing the thermodynamic potential used for the description of piezoelectricity,

$$\Phi_G = \frac{1}{2\varepsilon_0 \chi} P^2 + \frac{c}{2} u^2 - PE - fP\nabla u - u\sigma - \vartheta Pu, \quad (2)$$

where $c$ is an elastic constant, $\chi$ is the susceptibility, $P$ is polarization, $u$ is strain, $\sigma$ is stress, $\vartheta$ is the piezoelectric coefficient, and $f$ is a flexoelectric coupling coefficient (in units of volts). Minimizing the potential relative to polarization i.e. setting $\partial \Phi_G / \partial P = 0$

$$\boldsymbol{P} = \varepsilon_0 \chi \boldsymbol{E} + \varepsilon_0 \chi \vartheta u + \varepsilon_0 \chi f \nabla u \quad (3)$$

The second term in (3) is a piezoelectric term and $d = \varepsilon_0 \chi \vartheta$ is piezoelectric coefficient (which we drop for centro-symmetric materials) and the last term is obviously a flexoelectric one. The flexoelectric coefficient, which is a fourth rank tensor, can be defined as $\mu = \chi f$, or

$$\mu_{klij} = \left( \frac{\partial P_i}{\partial (\partial u_{kl} / \partial x_j)} \right)_{E=0} \quad (4)$$

Let us now consider a material that is not piezoelectric but has third order optical polarizability $\chi^{(3)}$ and re-write the thermodynamic potential as

$$\Phi_G = \frac{1}{2\varepsilon_0 \left( \chi + \chi^{(3)} \boldsymbol{EE} \right)} P^2 + \frac{c}{2} u^2 - \boldsymbol{PE} - f\boldsymbol{P}\nabla u - u\sigma \quad (5)$$

Minimizing this potential immediately yields

$$\boldsymbol{P} = \varepsilon_0 \chi \boldsymbol{E} + \varepsilon_0 \chi^{(3)} \boldsymbol{EEE} + \varepsilon_0 \chi f \nabla u + \varepsilon_0 f \nabla u \chi^{(3)} \boldsymbol{EE} \quad (6)$$

The last term in (6) can be written as $\boldsymbol{P} = \varepsilon_0 \chi_{SGI}^{(2)} \boldsymbol{EE}$, where SGI $\chi^{(2)}$ is $\chi_{SGI}^{(2)} = \psi \nabla u$, and where we introduced **flexo-nonlinear coefficient** as a sixth rank tensor

$$\psi_{mnklij} = \left( \frac{\partial \chi_{mni}^{(2)}}{\partial (\partial u_{kl} / \partial x_j)} \right)_{E=0} = \chi_{mnij}^{(3)} f_{kl} \quad (7)$$

It is easy to see [24,25] by comparing (6) with (1) that the term $f\nabla u$ plays the role of the effective field $E_{eff(\nabla u)}$ and once $f$ is known SGI $\chi^{(2)}$ can be determined as well.

Indeed, the flexo-coupling coefficient $f$ is easy to estimate from rather general considerations as has been first done by Kogan [34]. He had considered a unit cell of material, say a cube with side $a$ of about few angstroms. When a huge strain gradient is on the scale of $\nabla u \sim 1/a$ is applied the cell becomes completely distorted. The polarization change is on the scale of $\delta P = ea/a^3 \approx e/a^2$ while the change in the energy density is on the scale of $\delta \Phi_G \sim (e^2 / 4\pi\varepsilon_0 a) / a^3$. Now, according to (2) $\delta \Phi_G = f(\delta P)(\partial u / \partial x)$ and one obtains $f \sim e/4\pi\varepsilon_0 a \sim 1-10V$. More refined calculations [37-40] and numerous measurements of flexoelectricity in single crystals [40-41] confirm this simple result and all yield $f$ in that range. It follows then that for strain gradients of $\nabla u \sim 10^4 m^{-1}$ that can be realistically formed, the value of the effective field is only about $10^5 V/m$ in accordance to the analysis performed by us in the previous section. For a typical material with $\chi^{(3)} \sim 10^{-19} m^2/V^2$ (similar to Si), the **flexo-nonlinear coefficient** $\psi$ does not exceed $10^{-18} m^2/V$, and SGI-$\chi^{(2)}$ should remain much less than 0.1pm/V, in agreement with our estimate in the previous section. Again, this value clearly disagrees with the body of evidence indicating that the actual SGI-$\chi^{(2)}$ can be at least three to four orders of magnitude higher the that.

## 4 ALTERNATIVE MECHANISM OF STRAIN-GRADIENT INDUCED X[(2)]

The connection between SGI-χ[(2)] and flexoelectricity not only establishes the fact that in single crystal materials both phenomena are very weak, but also help us to establish an understanding of why and in which materials SGI-χ[(2)] can be high. While the magnitude of the flexoelectric effect in all single crystal materials is indeed low, the situation is different in perovskite ceramics [42-45] where the magnitude of the flexoelectric coefficient $f$ in excess of 1000 has been measured. In liquid crystals, the flexoelectric coefficients are even higher than that [46, 47]. To comprehend the origin of this large SGI polarizability, one must recall that in both perovskite ceramics and liquid crystals the individual bonds (individual molecules) already possess large electric dipoles, but, due to their random orientation, a macroscopic polarization is absent. The strain gradient, however, provides a preferential direction along which the bonds (molecules) tend to align. As mentioned above, a strain gradient acts in a fashion similar to the electric field, and the net result is "poling" of the material, conceptually similar to poling by the electric field in perovskites (LiNbO3) [48] or polymers [49]. It is important that no complete orientation is required – as long as the average projection of the individual dipole on the axis of strain gradient is distinct from zero, a reasonably large macroscopic polarization would result, as indeed the case in liquid crystals and ceramics subject to strain gradients.

Considering the similarity between flexoelectricity and SGI-$\chi^{(2)}$ established here, and the fact that SGI-$\chi^{(2)}$ is always associated with the presence of SiN it is then reasonable to make a conjecture that the origin of the SGI-$\chi^{(2)}$ is strain-gradient-induced "poling" of individual bonds in SiN ceramics. To estimate the magnitude of the "poling", consider Fig. 2a in which the projection of the 3D lattice of SiN onto the 2D plane is shown schematically. The SiN bonds are oriented essentially randomly in space. Each (n-th) bond has a second order polarizability with magnitude $\beta_{SiN}^{(2)}$ comparable to that of GaAs and directed along unit vector $\boldsymbol{b}_n$. Real SiN material contains excessive Si as well as hydrogen ions [50], but their introduction does not change the overall picture, other than making the structure less rigid and making the bond easier to turn. When the electric field is applied along the z-axis, a second order dipole moment $\boldsymbol{p}_n = \beta_{SiN}^{(2)} \left(\boldsymbol{b}_n \cdot \boldsymbol{E}\right)^2 \boldsymbol{b}_n = \beta_{SiN}^{(2)} E^2 \cos^2 \theta_n \boldsymbol{b}_n$ where $\theta_n$ is an angle between the n-th bond and vertical axis, is induced in each bond. The second order polarization is obtained by summation over n in a unit of volume

$$\boldsymbol{P}^{(2)} = V^{-1} \sum_n^N \boldsymbol{p}_n^{(2)} = 4 N_{Si} \beta_{SiN}^{(2)} E^2 \left\langle \cos^2 \theta_n \boldsymbol{b}_n \right\rangle_n = 0 \quad (8)$$

where $N_{Si}$ is the density of Si atoms and the brackets indicates averaging over the polar angle $\theta_n$ and the azimuthal angle $\varphi_n$. For the most relevant projection onto the z-axis it easy to see that $P_z^{(2)} \sim \left\langle \cos^3 \theta_n \right\rangle_{\theta_n} = 0$. Note that individual tetragons possess a second order polarizability, but since they are all randomly oriented the total polarization is zero.

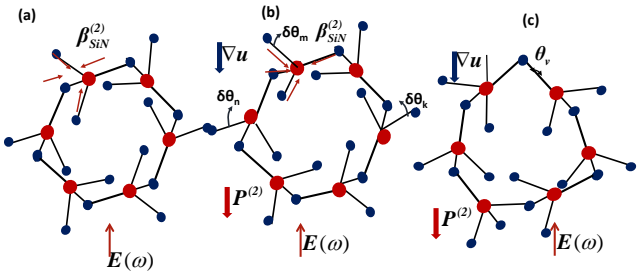

Fig. 2. (a) SiN lattice in the absence of a strain gradient (b) In the presence of a strain gradient individual Si-N rotate towards alignment with the strain gradient. (c) In the presence of a strain gradient rigid SiN4 tetragons rotate to align with the strain gradient

Now, when the strain gradient is applied one can consider two possibilities. First, , the tetragonal bonds get distorted and the angle between the bonds equal to 108.9 degrees is no longer maintained, as shown in Fig. 2b. As mentioned in the literature on the flexoelectric effect [24,25], the strain gradient acts as an effective electric field. This field, as has been shown above, is not sufficient to cause the displacement of charge density inside the bonds that would be commensurate with the observed values of nonlinear susceptibilities. However, this field may be sufficient to cause small rotation of bonds, particularly during the growth/deposition process. Since Si and N atoms have different charges (and also masses) the bonds would rotate to align in a vertical direction. The vertical effective force exerts a torque proportional to $\boldsymbol{b}_n \times \boldsymbol{z} = \sin \theta_n$, rotating each bond by the angle $\delta \theta_n = K \sin \theta_n$ where $K$ is a coefficient that depends on the conditions of growth and the strain gradient. The average angle of rotation is $\langle \delta \theta \rangle = K \langle \sin \theta_n \rangle = K \pi / 2$. Substituting it into (8) one obtains

$$\delta P_z^{(2)} = 4 N_{Si} \beta_{SiN}^{(2)} E^2 K \left\langle 3 \cos^2 \theta_n \sin^2 \theta_n \right\rangle_n = 4 N_{Si} \beta_{SiN}^{(2)} E^2 \frac{4}{5} K =$$
$$= 4 N_{Si} \frac{8}{5\pi} \beta_{SiN}^{(2)} \langle \delta \theta \rangle E^2 = \varepsilon_0 \chi_{SGI}^{(2)} E^2$$
(9)

We then can estimate the order of magnitude the SGI-$\chi^{(2)}$ in silicon nitride by comparing it with $\chi^{(2)}$ in GaAs which according to [18] is $\chi_{GaAs}^{(2)} = 4 N_{Ga} \left(\sqrt{3}/4\right) \varepsilon_0 \cdot \beta_{GaAs}^{(2)}$. The order of magnitude relation is

$$\frac{\chi_{SGI}^{(2)}}{\chi_{GaAs}^{(2)}} \approx 1.5 \frac{N_{Si}}{N_{Ga}} \frac{\beta_{SiN}^{(2)}}{\beta_{GaAs}^{(2)}} \langle \delta \theta \rangle \quad (10)$$

The nonlinear polarizabilities of SiN and GaAs bonds should be of comparable magnitude since on the one hand SiN has wider bandgap (and hence shorter bond length), but on the other hand the polarity is larger in SiN. In addition, the density of bonds in $Si_3N_4$ is higher. Therefore, for order-of magnitude estimate one can simply assume the term in front of $\langle \delta \theta \rangle$ in Eq. (10) equal to 1.5. Then, to obtain a SGI-$\chi^{(2)}$ equal to 10% of GaAs, i.e. on the scale of few tens of pm/V (comparable to measurements and to LiNbO3) the average bond rotation angle $\langle \delta \theta \rangle$ should be around 3-5 degrees. This number is relatively small and it is quite possible that in heavily hydrogenated $Si_3N_4$ this indeed happens during the growth process. In fact, during growth the individual bonds may behave in way similar to liquid crystal molecules that are known to align along the a strain gradient [31, 47].

Another possibility is that under a strain gradient the individual tetragons remain rigid, but they do rotate relative to each other in such a way that the probability of the vertical Si-N bond pointing upward is higher than probability of it pointing downward. This situation is shown in Fig. 2c and the bonds that are closest to being vertical are shown in thicker lines. Then, the nonlinear dipoles of individual tetragons no longer compensate each other and a nonzero SGI- χ(2) ensues. To estimate the nonlinear susceptibility we describe the orientation of n-th tetragon by the angle made by the most vertical of four bonds (shown by the thick line in Fig. 3c) with the vertical axis, $\theta_v$. Then, we assume that the probability of a given orientation can be written as $F(\theta_v) \sim 1 + \alpha \cos(\theta_v)$. It means that an "upward looking" orientation is $(1+\alpha)/(1-\alpha)$ times more probable than a "downward looking" orientation. Now, performing a summation of individual bond contributions, one can obtain the relation

$$\frac{\chi^{(2)}_{SGI}}{\chi^{(2)}_{GaAs}} \approx 0.7 \alpha \frac{N_{Si}}{N_{Ga}} \frac{\beta^{(2)}_{SiN}}{\beta^{(2)}_{GaAs}} \qquad (11)$$

This indicates that for $\alpha \sim .14$, i.e. a relatively small degree of "tetragon poling", the values of SGI- χ(2) reported in the literature will result.

Note that one of the possible mechanisms for a non-zero second order susceptibility in SiN invoked in the literature [21-23] has been the re-orientation of Si clusters with broken inversion symmetry [51], but no quantitative estimate of the effect had been given. Since each Si-N bond (or tetragon) already carries a large nonlinear polarizability that leads to a large χ(2) with a strain-gradient induced alignment, , it appears to us that it is entirely unnecessary to invoke Si clusters whose nonlinear polarizability is inherently small (since the bonds inside are homopolar).

Likewise, the large experimentally observed χ(2) cannot be explained by a surface nonlinearity, because surface χ(2) effects in a waveguide mode of effective width $w_{eff}$ is "diluted" roughly by a confinement factor $a_b / w_{eff} < 10^{-3}$ which would yield χ(2) on the scale of much less than 1pm/V. Only bulk effects can produce the measured values of χ(2) (~tens pm/V) observed in Refs. [6, 10, 12-14].

## 5. DISCUSSION AND CONCLUSIONS

In this work we have addressed the nature of the second-order nonlinear effects observed in strained silicon on silicon nitride and pure silicon nitride materials. By connecting the χ(2) induced by strain gradients to the flexoelectric effect and by developing our own effective field theory, we have shown that in the absence of heteropolar bonds, as is the case in Si, the magnitude of the SGI- χ(2) remains at least three orders of magnitude less than that observed experimentally. In this, our result strongly disagrees with those in Ref [16] where large cluster of homopolar bonds with no microscopic nonlinear polarizability somehow managed to produce a large macroscopic χ(2).

At the same time, when heteropolar bonds with large nonlinear polarizabilities are already present, as is the case with silicon nitride, just a small re-orientation of these bonds can produce a large χ(2). It is our conclusion that nonlinear and electro-optic effects observed in the strained Si waveguides, starting with [6], are occurring in the silicon nitride cladding. Our results are supported by the measurements performed in [10], in which χ(2) was mapped using THz generation and shown to originate from the cladding (Fig. 5b). Further work [52] has shown that effective χ(2) increases with the reduction of the waveguide dimensions, which can also be explained by the increased mode penetration into the silicon nitride cladding. It was further shown in [53] that the nonlinearity can be activated by the formation of Si-Br bonds during an HBr dry etching process, and we predict a similar effect for SiN bonds. Finally, the fact that χ(2) has been observed in SiN waveguides that are naturally strained [19-23] also serves to confirm our explication.

It is important that in our view the SGI- χ(2) owes its existence to the amorphous, or rather ceramic structure of SiN, which, coupled with the large number of defects due to hydrogenation, allows the bonds to be poled by the amount that is small but sufficient to explain the observed SGI- χ(2). This re-orientation would be highly improbable in a rigid single crystal lattice, consistent with the fact that the flexoelectric effect becomes substantial only in amorphous materials.

To conclude, we reiterate that electro-refractive modulators in CMOS-compatible materials that lack free-carriers can be faster, more efficient, and more linear than current free-carrier based devices. In addition, a second-order susceptibility in these materials is critical for efficient frequency converters and parametric oscillators. Together with additional experiments that focus on strain gradients in SiN, we believe that our explication will lay the foundation for these improved silicon-based waveguide devices. Such waveguides could be based on either silicon cores with strained SiN claddings or directly strained SiN cores.

**Acknowledgement**

Fruitful discussions with B. Jalali of UCLA are acknowledged. JBK is grateful to the ONR Summer Faculty Research Program for providing him with an opportunity to work on this project.

**References**

1. L. Pavesi, L. and D.J. Lockwood, *Silicon Photonics* (Springer, Berlin, 2004).
2. Q Xu, B Schmidt, S Pradhan, M Lipson, "Micrometre-scale silicon electro-optic modulator", Nature **435**(7040), 325–327 (2005).
3. L. Liao, A. Liu, D. Rubin, J. Basak, Y. Chetrit, H. Nguyen, R. Cohen, N. Izhaky, and M. Paniccia, "40 Gbit/s silicon optical modulator for high-speed applications," Electron. Lett. **43**(22), 1196–1197 (2007).
4. C. Koos , P. Vorreau , T. Vallaitis , P. Dumon , W. Bogaerts , R. Baets ,B. Esembeson , I. Biaggio , T. Michinobu , F. Diederich , W. Freude ,J. Leuthold , "All-optical high-speed signal processing with silicon–organic hybrid slot waveguides", Nat. Photonics, **3** , 216 (2009)
5. H-W Chen, J. D. Peters, and J. E. Bowers,"Forty Gb/s hybrid silicon Mach-Zehnder modulator with low chirp", Opt. Express, **19**,1455 (2011)

6. R. S. Jacobsen, K. N. Andersen, P. I. Borel, J. Fage-Pedersen, L. H. Frandsen, O. Hansen, M. Kristensen, A. V. Lavrinenko, G. Moulin, H. Ou, C. Peucheret, B.Zsigri, and A. Bjarklev, "Strained silicon as a new electro-optic material" Nature **441**, 199 (2006).
7. N. K. Hon, K. K. Tsia, D. R. Solli, and B. Jalali, Appl. Phys. Lett. "Peridically poled Ssilicon", **94**, 091116 (2009).
8. S. V. Govorkov, V. I. Emel'yanov, N. I. Koroteev, G. I.Petrov, I. L. Shumay, and V. V. Ya kovlev, "Inhomogeneous deformation of silicon surface layers probed by second-harmonic generation in reflection", J. Opt. Soc.Am. B **6**, 1117 (1989).
9. J. Y. Huang, "Probing Inhomogeneous Lattice Deformation at Interface of *Si*(111)/*SiO*$_2$ by Optical Second-Harmonic Reflection and Raman Spectroscopy", Jpn. J. Appl. Phys. **33**, 3878 (1994).
10. B. Chmielak, M. Waldow, C. Matheisen, C. Ripperda, J. Bolten, T. Wahlbrink, M. Nagel, F. Merget, and H.Kurz, "Pockels effect based fully integrated, strained silicon electro-optic modulator," Opt. Express **19**(18),17212–17219 (2011).
11. I. Avrutsky, and R. Soref "Phase-matched sum frequency generation in strained silicon waveguides using their second order nonlinear optical susceptibility", Opt. Express **19**(22), 21707 (2011)
12. M. W. Puckett, J. S. T. Smalley, M. Abashin, A. Grieco, and Y. Fainman, Tensor of the second-order nonlinear susceptibilityin asymmetrically strained silicon waveguides: analysis and experimental validation", Opt. Lett, **39**, 1693 (2014)
13. C. Schriever , C. Bohley , R. B. Wehrspohn , Strain dependence of second-harmonic generation in silicon ,Opt. Lett., **35**, 273 (2010) .
14. C. Schriever , F. Bianco , M. Cazzanelli , M.Ghulinyan ,C. Eisenschmidt , J. de Boor , A. Schmid , J. Heitmann ,L.Pavesi , and J.Schilling , "Second-Order Optical Nonlinearity in Silicon Waveguides", **3**, 129–136 (2015)
15. N. K. Hon, K. K. Tsia, D. R. Solli, B. Jalali, and J. B. Khurgin, "Stress-induced χ(2) in silicon – comparison between theoretical and experimental values," in IEEE 6th International Conference on Group IV Photonics, San Francisco, CA (9-11 September 2009).
16. M. Cazzanelli, F. Bianco. E. Borga, G. Pucker, M. Ghulinyan, E. Degoli, E. Luppi, V. Véniard, S. Ossicini, D. Modotto, S.Wabnitz, R. Pierobon and L. Pavesi, "Second-harmonic generation in silicon waveguides strained by silicon nitride", Nature Materials, **11**, 148 (2012)
17. S. Sharif Azadeh, F. Merget, M. P. Nezhad, and J. Witzens, "On the measurement of the Pockels effect in strained silicon", Opt. Lett., **40**, 1847 (2015)
18. B. F. Levine, "Bond-charge calculation of nonlinear optical susceptibilities for various crystal structures". Phys. Rev. B **7**, 2600 (1973)
19. T. Ning, H. Pietarinen, O. Hyvrinen, J. Simonen, G. Genty,and M. Kauranen, "Strong second-harmonic generation in silicon nitride films Appl. Phys. Lett. **100**, 161902 (2012).
20. J. S. Levy , M. A. Foster , A. L. Gaeta , M. Lipson , "harmonic generation in silicon nitride ring resonators", Opt. Express, **19** , 11415 (2011**)**
21. S. Miller, Y-H D Lee, J Cardenas, A. L .Gaeta, M. Lipson, "Electro-Optic Effect in Silicon Nitride", CLEO 2015, SF1G.4 (2015)
22. A. Kitao, K. Imakita, I. Kawamura and M. Fujii An investigation into second harmonic generation by Si-rich SiN$_x$ thin films deposited by RF sputtering over a wide range of Si concentrations J. Phys. D: Appl. Phys. **47** 215101(2014)
23. E. F. Pecora, A. Capretti, G. Miano, and L. Dal Negro, Generation of second harmonic radiation from sub-stoichiometric silicon nitride thin films Appl. Phys. Lett. **102**, 141114 (2013)
24. P. Zubko P, O. Catalan G and A. K. Tagantsev "Flexoelectric effect in solids" Annu. Rev. Mater. Res. **43** 387–421 ( 2013)
25. P V Yudin and A K Tagantsev Fundamentals of flexoelectricity in solids Nanotechnology 24 432001 (2013)
26. J. Phillips, Dielectric definition of electronegativity. *Physical Review Letters* **20**, 550 (1968).
27. J. A. Van Vechten, "Quantum dielectric theory of electronegativity in covalent systems". I. Electronic dielectric constant. *Physical Review* **182**, 891 (1969).
28. J. Van Vechten, "Quantum dielectric theory of electronegativity in covalent systems. II. Ionization potentials and interband transition energies". *Physical Review* **187**, 1007 (1969).
29. N. Christensen, S. Satpathy, Z. Pawlowska, "Bonding and ionicity in semiconductors". *Physical Review B* **36**, 1032 (1987).
30. H. P. Wagner and M. Kuhnelt, W. Langbein and J. M. Hvam, "Dispersion of the second-order nonlinear susceptibility in ZnTe, ZnSe, and ZnS", Phys. Rev. B, **58**, 10494 (1998)
31. M,. Dinu, F. Quochi, and H. Garcia, "Third Order nonlinearities in silicon at telecom wavelengths", Appl.Phys.Lett, **82**, 2954(2003)
32. R. B. Meyer "Piezoelectric effects in liquid crystals" Phys. Rev. Lett. **22** 918–21 (1969)
33. V. S. Mashkevich , K B Tolpygo Electrical, optical and elastic properties of diamond type crystals. 1 Sov.Phys.—JETP **5** 435–9 (1957)
34. K.B. Tolpygo "Long wavelength oscillations of diamond-type crystals including long range forces Sov. Phys Solid State **4** 1297–305 (1963)
35. S M Kogan, "Piezoelectric effect during inhomogeneous deformation and acoustic scattering of carriers in crystals" Sov. Phys. Solid State **5** 2069–70 (1964)
36. P. Harris, "Mechanism for the shock polarization of dielectrics" J. Appl. Phys. **36** 739–41 (1965)
37. R. Maranganti , P. Sharma P. "Atomistic determination of flexoelectric properties of crystalline dielectrics". Phys. Rev. B **80**:054109, (2009).
38. J. Hong , G. Catalan , J. F. Scott , E. Artacho . "The flexoelectricity of barium and strontium titanates from first principles" J. Phys. Condens. Matter **22**(11):112201 (2010)
39. I. Ponomareva , K. Tagantsev L. Bellaiche "Finite-temperature flexoelectricity in ferroelectric thin films from first principles". Phys. Rev. B **85**:104101(2012)
40. P. Zubko , G. Catalan G, A. Buckley, P. R. L. Welche and J F Scott "Strain-gradient-induced polarization in SrTiO3 single crystals", Phys. Rev. Lett. **99** 167601(2007)
41. E. Farhi , A. K. Tagantsve ,R.Currat , B. Hehlen E. Courtens and L. A. Boatner "Low energy phonon spectrum and its parameterization in pure KTaO3 below 80 k", Eur. Phys. J.B **15** 615–23 (2000)
42. L. Cross "Flexoelectric effects: charge separation in insulating solids subjected to elastic strain gradients J. Mater. Sci. **41** 53–63 (2006)

43. W. Ma and L. E. Cross, Strain-gradient-induced electric polarization in lead zirconate titanate ceramics", Appl. Phys. Lett. **82** 3293–5 (2003)
44. W. Ma and L. E. Cross, "Flexoelectricity of barium titanate", Appl. Phys. Lett. **88** 232902 (2006)
45. D. Lee, A. Yoon, S.Y. Jang, J.-G. Yoon, J.-S. Chung, M. Kim, J. F. Scott,5and T.W. Noh, "Giant Flexoelectric Effect in Ferroelectric Epitaxial Thin Films", Phys Rev. Lett, **107**, 057602 (2011)
46. H. J. Coles and M. N. Pivnenko, "Liquid crystal 'blue phases' with a wide temperature range", Nature **436**, 997 (2005).
47. J. S. Patel, and R. B. Meyer, "Flexoelectric electro-optics of a cholesteric liquidcrystal". Phys. Rev. Lett. **58**, 1538–-1540 (1987).
48. E.J. Lim, M.M. Fejer, R.L. Byer, W.J. Kozlovski, Blue light generation by frequency doubling in periodically poled lithium niobate channel waveguide, Electronics Letters, **25**,(11),. 731 – 732 (1989)
49. K. D. Singer, M. G. Kuzyk, W. R. Holland, J. E. Sohn, S. J. Lalama, R. B. Comizzoli, H. E. Katz and M. L. Schilling "Electro-optic phase modulation and optical second-harmonic generation in corona-poled polymer films", Appl. Phys. Lett. **53**, 1800 (1988)
50. M. Blech, A. Laades, C. Ronning, B. Schroter, C. Borschel, D. Rzesanke, A. Lawerenz, "Detailed Study of PECVD Silicon Nitride And Correlation of Various Characterization Techniques", Proceedings of the 24th European Photovoltaic Solar Energy Conference and Exhibition, Hamburg, Germany, 2009, 507-511.
51. T T, Rantala , M. I. Stockman, D. A. Jelski and T. F. George Linear and nonlinear optical properties of small silicon clusters J. Chem. Phys. **93** 7427–38 (1990)
52. B. Chmielak, C. Matheisen, C. Ripperda, J. Bolten, T. Wahlbrink, M. Waldow, and H. Kurz, "Investigation of local strain distribution and linear electro-optic effect in strained silicon waveguides," Opt. Express **21**, 25324-25332 (2013)
53. C. Matheisen, M. Waldow, B. Chmielak, S. Sawallich, T. Wahlbrink, J. Bolten, M. Nagel, and H. Kurz, "Electro-optic light modulation and THz generation in locally plasma-activated silicon nanophotonic devices," Opt. Express **22,** 5252-5259 (2014).